\documentclass[aps,prb,twocolumn,raggedbottom,showpacs,floatfix,superscriptaddress]{revtex4}

\usepackage{amsmath,amsfonts,amssymb,amsthm,graphics}
\usepackage[next]{inputenc}
\usepackage[dvips]{epsfig}
\usepackage[colorlinks=true,citecolor=blue,linkcolor=blue]{hyperref}
\usepackage{bbm}
\usepackage{bbold}
\usepackage{booktabs}
\usepackage{multirow}
\usepackage{hhline}
\usepackage{amssymb}
\usepackage{comment}

\usepackage{bm}
\usepackage{color}
\usepackage{graphicx,latexsym}
\usepackage{epstopdf}

\usepackage{esvect}
\usepackage{fixmath}

\def\be{\begin{equation}}
\def\ee{\end{equation}}

\def\bi{\begin{itemize}}
\def\ei{\end{itemize}}
\def\bn{\begin{enumerate}}
\def\en{\end{enumerate}}
\def\bea{\begin{eqnarray}}
\def\eea{\end{eqnarray}}
\newcommand{\bpm}{\begin{pmatrix}}
\newcommand{\epm}{\end{pmatrix}}

\def\ba{\begin{array}}
\def\ea{\end{array}}
\def\bd{\begin{displaymath}}
\def\ed{\end{displaymath}}

\renewcommand{\imath}{\hspace{1pt}\mathrm{i}\hspace{1pt}}
\def\phs{\mathcal{C}}
\def\sor{\mathcal{U}_1}

\begin{document}
\title{Deformation and Stability of Surface States in Dirac Semimetals}
\author{Mehdi Kargarian}
\affiliation{Department of Physics, Sharif University of Technology, Tehran 14588-89694, Iran}
\author{Yuan-Ming Lu}
\affiliation{Department of Physics, The Ohio State University, Columbus, OH 43212, USA}
\author{Mohit Randeria}
\affiliation{Department of Physics, The Ohio State University, Columbus, OH 43212, USA}

\begin{abstract}
The unusual surface states of topological semimetals have attracted a lot of attention. Recently, we showed [PNAS 113, 8648 (2016)] that for a Dirac semimetal (DSM) arising from band-inversion, such as Na$_3$Bi and Cd$_3$As$_2$, the expected double Fermi arcs on the surface are not topologically protected. Quite generally, the arcs deform into states similar to those on the surface of a strong topological insulator. Here we address two questions related to deformation and stability of surface states in DSMs. First, we discuss why certain perturbations, no matter how large, are unable to destroy the double Fermi arcs. We show that this is related to certain extra (particle-hole) symmetry, which is non-generic in materials. Second, we discuss situations in which the surface states are completely destroyed without breaking any symmetry or impacting the bulk Dirac nodes. We are not aware of any experimental or density functional theory (DFT) candidates for a material which is a bulk DSM without any surface states, but our results clearly show that this is possible.


\end{abstract}
\date{\today}
\pacs{73.43.-f, 73.20.-r, 73.43.Cd, 71.20.-b}

\maketitle


\section{Introduction \label{intro}}

In the last decade, condensed matter physics witnessed a large class of novel electronic materials characterized by their topological properties, such as quantized bulk response functions and robust surface states. Experimental discovery of topological insulators protected by time reversal symmetry \cite{Hasan:rmp10,Qi:rmp11,HasanMoore:Ann11} greatly advanced our understanding of quantum states with nontrivial topology, which are beyond the description of symmetry breaking and local order parameters in the traditional Landau paradigm. Moreover, these topological properties also apply to metallic systems, where the gapless bulk band structures are featured by symmetry protected point or line nodes\cite{volovik:book,Horava2005,vishwanath:prb11,burkov:prb11,Zhao2013,Matsuura2013}. For the former case, the conduction and valence bands intersect at isolated points at the Fermi level, giving rise to low energy electron behaviors described by relativistic Dirac equations\cite{vafek:annal14,Hosur:crp13,Hasan2017,Yan2017,Armitage2017}. One outstanding example is the three-dimensional (3D) Dirac material, a 3D analogue of graphene, featured by nodal Dirac points at the Fermi level where the bands disperse linearly in all directions\cite{zaheer:prl12,wang:prb12,wang:prb13}. After the theory predictions\cite{wang:prb12,wang:prb13}, Dirac semimetals have been experimentally discovered in Na$_3$Bi and Cd$_3$As$_2$ by angle-resolved photoemission spectroscopy (ARPES)\cite{Li:science14,Ong:epl16,Yi:Sciencereport14,Neupane:ncomm14}, scanning tunneling spectroscopy (STM)\cite{Jeon:nmat14} and magneto-transport\cite{Analytis:Nat.2016} measurements.

\begin{figure*}[!htb]
\includegraphics[width=17cm]{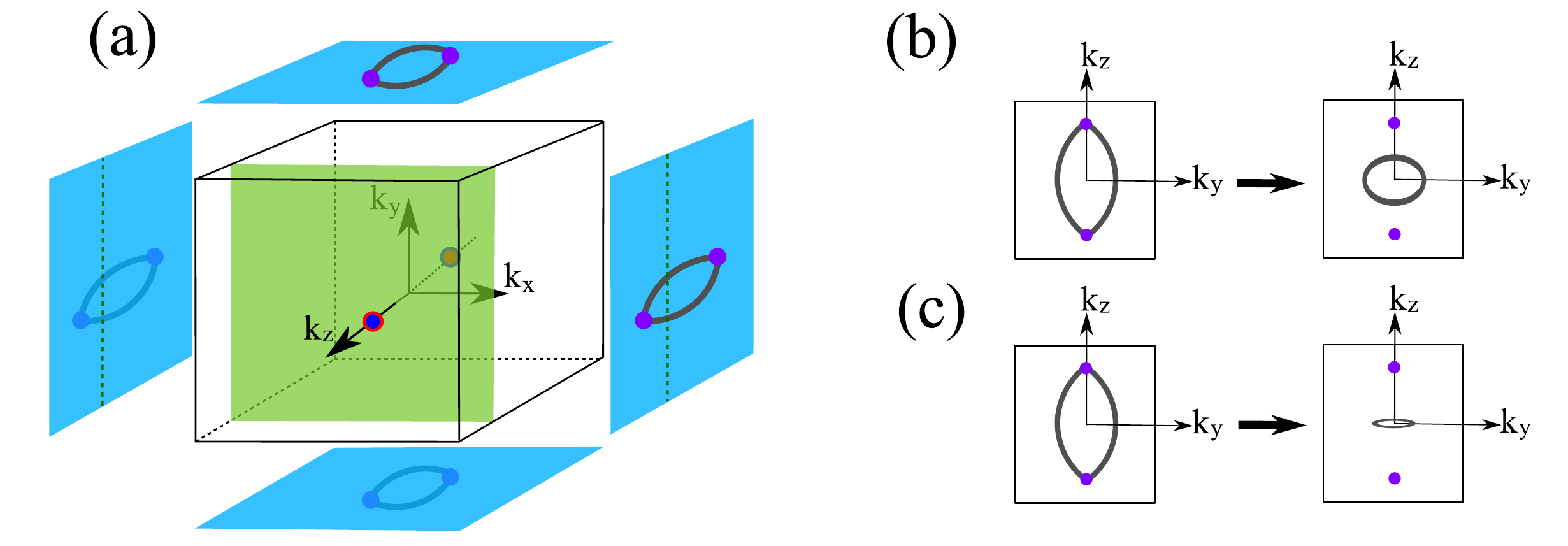}
\caption{(Color online) (a) Schematic {\bf k}-space picture showing the location of Dirac nodes each superimposed of two Weyl nodes (red and blue) along $k_z$ axis in bulk Brillouin zone (BZ)
and the possible double Fermi arcs on the surface BZ's, shown as blue squares. Note that surfaces perpendicular to $z$ axis have no arcs. A 2D slice of BZ perpendicular to $k_z$ axis is shown as green square projected to green dashed lines on side surfaces, where we only showed for (100) side surface.
(b, c) Continous deformation of double Fermi arcs on (100) side surface by coupling zero modes on Fermi arcs. Perturbations near the projected node can deform the Fermi arcs to a closed contour in right surface BZ in (b). Most parts of latter can be gapped out by coupling more states on double Fermi arcs and leaves behind only gapless helical modes concentrated around the center of surface BZ in (c).} \label{fig:arcs_fermisurfaces}
\end{figure*}

With both time-reversal and inversion symmetries in a Dirac semimetal (DSM), each Dirac node is described by a four-component Dirac equation, which can be viewed as two degenerate Weyl fermions with opposite chiralities. Extra crystalline symmetires are required to forbid the chirality mixing of these two Weyl fermions and to stablize a DSM\cite{zaheer:prl12,yang:ncomm2014}. For example, Na$_3$Bi and Cd$_3$As$_2$ belong to a class of DSMs, which originate from band inversions\cite{Armitage2017} in a way similar to 3D topological insulators (TIs). In a 3d TI, the crossings between valence and conduction bands from the band inversion are all gapped out, resulting in an insulating ground state. On the other hand, in a 3d \emph{DSM from band inversions}, these band crossings remain gapless along certain high-symmetry axis (associated with a crystalline symmetry) and hence lead to a semimetal.

If separated in momentum-space, bulk Weyl nodes generally lead to open Fermi arcs on surfaces of system \cite{vishwanath:prb11} connecting two Weyl nodes with opposite chiralities, as observed in non-centrosymmetric Weyl semimetal material TaAs \citep{Xu:science2015_weyl,weng:prx15,lv:prx15,Lv:nature15}. Therefore, when a pair of time-reversal-related Dirac nodes are located off the time reversal invariant momenta (TRIM) in the Brillouin zone (BZ), a pair of Fermi arcs (``double Fermi arcs'') are naively expected to appear on certain surfaces of the system\cite{wang:prb12,wang:prb13, Gorbar:prb15} as shown schematically in Fig. \ref{fig:arcs_fermisurfaces}(a). Unlike a Weyl semimetal stable against any perturbations\cite{vishwanath:prb11}, a DSM requires the protection of extra crystalline symmetries such as a $(2n)$-fold rotational symmetry in Na$_3$Bi and Cd$_3$As$_2$\cite{wang:prb12,wang:prb13,yang:ncomm2014}, which are explicitly broken on open surfaces of a sample. Therefore it is not obvious whether it is valid or not to apply the usual bulk-boundary correspondence of topological states to a DSM. Motivated by the large and increasing pool of DSM materials, it is important to understand the stability of surface states in a 3D DSM.

In a previous work\cite{KARGARIAN2016}, we showed that contrary to naive expectations\cite{Potter2014}, the surface double Fermi arcs in a DSM are unstable against certain perturbations. They can be continuously deformed into a closed pocket similar to a 3d TI surface (see FIG. \ref{fig:arcs_fermisurfaces}(b)-(c)), without any symmetry breaking or bulk phase transitions, which can affect the surface optical properties\cite{shi:prb2017}. Regarding the stability and deformation of DSM surface states, two issues remained unresolved in our previous work. The first issue involves the different perturbations to the ${\bf k}\cdot{\bf p}$ Hamiltonian of DSM materials Na$_3$Bi and Cd$_3$As$_2$. While some perturbations can deform the surface double fermi arcs into a closed pocket, some other perturbations can never destroy the double fermi arcs no matter how strong they are. While both types of perturbations keep the bulk Dirac nodes stable, what causes such a dramatic difference in their effects on the surface states? The second issue involves the stability of surface states in a general DSM from band inversions. While Na$_3$Bi and and Cd$_3$As$_2$ do host side surface states that cannot be removed by any symmetric perturbations, is this always true for a DSM from band inversions? Can there be a DSM with no robust surface states on any open surfaces?

In this work we resolve these two issues on the stability and deformation of DSM surface states. The paper is organized as follows. In section \ref{sec:numerics} we set up the minimal model and associated symmetries for a DSM from band inversions, adopted from the ${\bf k}\cdot{\bf p}$ Hamiltonian of Cd$_3$As$_2$ and Na$_3$Bi. In section \ref{sec:surface states deformation} we resolved the first issue, by showing that a non-generic particle-hole symmetry (not present in materials) in the minimal model is responsible for the stability of double fermi arcs under certain perturbations. In section \ref{sec:chirality mixing} we further show that those perturbations breaking this non-generic particle-hole symmetry can effectively mix the two Weyl fermions with opposite chiralities away from the bulk Dirac nodes, and therefore deform the double fermi arcs into a closed pocket on the surface. In section \ref{sec:dsm with no surface}, we resolve the second issue with a positive answer, by explicitly constructing an 8-band DSM model with robust bulk Dirac nodes and no surface states at all. We show that DSMs with and without surface states can both be understood in terms of a bulk $Z_2$ topological invariant, which can be defined for DSMs from band inversions despite the absence of a bulk gap. Finally we conclude in section \ref{sec:conclude}.

\section{The minimal model of Dirac semimetals and its symmetries}\label{sec:numerics}
We begin our discussion by introducing a model Hamiltonian describing the essential feature of Dirac semimetals, which also applies to the materials \cite{wang:prb12,wang:prb13}. The simplest model would be a 4-band model with $n=4$-fold rotational symmetry along $\hat z$-axis. Despite being simple, this model captures the band inversion near the $\Gamma$ point of BZ of Cd$_3$As$_2$ \cite{wang:prb13}.

The 4-band model describing Dirac semimetal Cd$_3$As$_2$ is given as follows:
\bea \label{H4} H(\textbf{k})&=&\varepsilon_\textbf{k}+\Big[t(\cos k_x+\cos k_y-2)+t_{z}(\cos k_z-b)\Big]\tau_{z}\nonumber \\
&&+\lambda \sin k_x\sigma_{x}\tau_{x}+\lambda \sin k_y\sigma_{y}\tau_{x}, \eea
where $\boldsymbol{\sigma}$ and $\boldsymbol{\tau}$ are Pauli matrices for spin and orbital indices. $t$($t_z$) is the in-plane (out of plane) hopping amplitude and $\lambda$ represents spin-orbit coupling (SOC) and we choose $b=\cos Q$. The dispersive term $\varepsilon_\textbf{k}=\varepsilon_{-{\bf k}}$ should be consistent with all symmetries of the system as follows and vanish at the two Dirac nodes.
For example two convenient choices are $\varepsilon_\textbf{k}=t_{1}(\cos k_{z}-\cos Q)+t_{2}(\cos k_{x}+\cos k_{y}-2)$ or $\varepsilon_\textbf{k}=t_{1}(\cos k_{z}-\cos Q)(\cos k_{x}+\cos k_{y})$. Adding a dispersive term $\varepsilon_{\bf k}\neq0$ makes the spectrum asymmetric and will provide curvature to the double Fermi arcs as shown in Fig. \ref{fig:arcs_fermisurfaces}. This term $\varepsilon_{\textbf{k}}\neq0$ neither moves the location of Dirac nodes in the BZ nor shift them in energy, so the two Dirac nodes remain as the only gap-closing points in the bulk.

In the long-wavelength limit, this model \cite{chiu:conference} is equivalent to the well-known effective $\mathbf{k}\cdot \mathbf{p}$ Hamiltonian around $\Gamma$ point for Dirac semimetal materials\cite{wang:prb12,wang:prb13}, up to a unitary transformation in the fermion basis. The spectrum is gapped over the entire Brillouin zone (BZ) except at two discrete points ${\bf k}=(0,0,\pm Q)$, where the low energy spectrum is described by linearly-dispersing Dirac fermions.

Clearly the Hamiltonian (\ref{H4}) preserves time-reversal $\Theta=\imath\sigma_{y}\cdot\mathcal{K}$, inversion $I$, two-fold rotation $C_{2,x(y)}$ and $n$-fold rotation $C_{n,z}$ ($n=4$ for Cd$_3$As$_2$ and $n=6$ for Na$_3$Bi) symmetries as follows.
\bea \label{symmetry1}&&\Theta^{-1} H(\textbf{k})\Theta=H(-\textbf{k}),~U_{I}^{-1}H(\textbf{k})U_{I}=H(-\textbf{k}),\\ \label{symmetry2}
&&U_{c_{2,x}}^{-1} H(k_x,k_y,k_z) U_{c_{2,x}}=H(k_x,-k_y,-k_z),\\ \label{symmetry3}
&&U_{c_{2,y}}^{-1} H(k_x,k_y,k_z) U_{c_{2,y}}=H(-k_x,k_y,-k_z),\\ \label{symmetry4}
&&U_{c_{n,z}}^{-1} H(k_{\pm},k_z) U_{c_{n,z}}=H(e^{\pm \imath\frac{2\pi}n}k_{\pm},k_z), \eea
where $k_{\pm}=k_x\pm \imath k_y$. The operators $U$'s describe the matrix representations of symmetry operation in spin and orbital basis. As written the Hamiltonian in (\ref{H4}) in this form, the symmetry operators should be as follows: inversion $U_{I}=\tau_{z}$, two-fold rotation $U_{c_{2,x(y)}}=\imath\sigma_{x(y)}$ and $n$-fold rotation $U_{c_{n,z}}=e^{\imath\pi\sigma_{z}(1-2\tau_z)/n}$. Note that this model also has mirror reflection symmetries with respect to major planes:
\bea U_{R_{i}}=U_{c_{2,i}} U_{I}=\imath\sigma_i\tau_z:~~~k_{i}\rightarrow -k_{i},\eea
where $i=x,y,z$.

In addition to the time reversal and crystal symmetries that exists in DSM material Cd$_3$As$_2$ as summarized above, the simplified 4-band minimal model (\ref{H4}) also exhibits two extra symmetries:

(i) unitary $U(1)$ symmetry $\{\sor(\phi)|0\leq\phi<2\pi\}$
\bea\label{sym:u1}
\sor(\phi) H({\bf k})\sor^\dagger(\phi)=H({\bf k}),~~~\sor(\phi)=e^{\imath\sigma_z\tau_z\frac\phi2}
\eea
which generates a $U(1)$ group of combined spin and orbital rotation;

(ii) particle-hole symmetry $\phs$ if $\varepsilon_{\bf k}=0$ in minimal model (\ref{H4})
\bea\label{sym:phs}
\phs H({\bf k})\phs^{-1}=-H^T(-{\bf k}),~~~\phs=\imath\sigma_x\tau_x\cdot\mathcal{K}
\eea
under which the 4-component fermion $\psi_{\bf k}$ transform as
\bea
\phs \psi_{\bf k}\phs^{-1}=\imath\sigma_x\tau_x\psi^\ast_{-{\bf k}}
\eea
Notice that even if $\varepsilon_{\bf k}\neq0$ in model (\ref{H4}), the wave functions of model (\ref{H4}) still preserves particle-hole symmetry $\phs$, since $\varepsilon_{\bf k}\neq0$ will only change the spectrum but not eigenstates of (\ref{H4}). As a result, particle-hole symmetry is effectively valid even if $\varepsilon_{\bf k}\neq0$.

In the presence of these two extra symmetries, the combination of time reversal $\Theta=\imath\sigma_{y}\cdot\mathcal{K}$ and particle-hole symmetry $\phs$ in (\ref{sym:phs}) becomes a (unitary) chiral symmetry \cite{chiu:conference}
\bea\label{sym:chiral}
\{H({\bf k}),\Theta\phs\}=0,~~~\Theta\phs=\imath\sigma_z\tau_x.
\eea
which leads to a particle-hole symmetric spectrum when $\varepsilon_{\bf k}=0$.

We emphasize that extra symmetries (\ref{sym:u1}) and (\ref{sym:phs}) are merely artifacts of the 4-band minimal model (\ref{H4}), but not an intrinsic property of the DSM materials. Same is true for the chiral symmetry (\ref{sym:chiral}). As will be shown later, these extra (unphysical) symmetries will play an important role in the formation of double-fermi-arc surface states of model (\ref{H4}). In a generic DSM material where these symmetries are absent, there won't be double fermi arcs on the surface of the DSM sample.

\section{Surface states}\label{sec:surface states deformation}
In this section we study the surface states localized on the [100] side surface of the system, which is parallel to the separation between two Dirac nodes. For simplicity we assume that the system fills the half space $x\geq0$, and the [100] side surface hosting the gapless surface states is the $y-z$ plane located at $x=0$. In order to get analytical expressions for surface states we linearize the model (\ref{H4}) around the line $\mathbf{k}=(0,0,k_z)$. We obtain

\bea \label{Hl} H_\text{Dirac}(\textbf{k})=m(k_z)\tau_{z}+\lambda k_x\sigma_{x}\tau_{x}+\lambda k_y\sigma_{y}\tau_{x}, \eea
where $m(k_z)=t_{z}(\cos k_z-b)$ and we ignore the terms arising from $\varepsilon_{\mathbf{k}}$ for simplicity as they do not affect the surface state wavefunctions. For a semi-infinite system described above, the momentum $\mathbf{k}$ is substituted by $\mathbf{k}=(-\imath\partial_{x},k_{y},k_{z})$. Therefore, the model (\ref{Hl}) for a semi-infinite systems reads as

 \bea \label{Hsemi} \tilde H_\text{Dirac}=m(k_z)\tau_{z}+\lambda\sigma_{x}\tau_{x} (-\imath\partial_{x})+\lambda k_y\sigma_{y}\tau_{x}. \eea

 For a fixed surface momentum $(k_y,k_z)$, the Hamiltonian (\ref{Hl}) effectively describes a one-dimensional (1d) model. Therefore, the zero-energy edge states is reduced to the well-known Jackiw-Rebi soliton \cite{Jackiw_Rebbi} localized around the mass domain wall of 1d Dirac Hamiltonians. We have to solve the eigenvalue problem
\bea
 \tilde H_\text{Dirac}|\psi_{k_y,k_z}(x)\rangle=E|\psi_{k_y,k_z}(x)\rangle
\eea
  The localized edge states should have the following form
 \bea |\psi_{k_y,k_z}(x)\rangle=|\chi\rangle e^{-x/\xi},  \eea
 where $\xi$ is a length scale over which the surface states decay into the bulk. After solving the eigenvalue problem, the zero-energy surface states have the following wavefunctions

 \bea \label{basis} |\psi^{1}(x)\rangle =\frac{Ae^{-x/\xi}}{\sqrt{2}}
 \begin{pmatrix}
 1 \\
  0\\
  0 \\
  \imath
 \end{pmatrix},~~|\psi^{2}(x)\rangle =\frac{Ae^{-x/\xi}}{\sqrt{2}}
 \begin{pmatrix}
 0 \\
  \imath\\
  1 \\
  0
 \end{pmatrix}\eea
where $\xi\sim\lambda/m(k_z)$ and $A$ is the normalization factor. It is clear that in the vicinity of gap closing in the bulk $k_{z}=\pm Q$ the surface states merge into the bulk states since $m(k_{z}\simeq \pm Q)\rightarrow 0$ and the localization length $\xi$ diverges.

Projecting the Hamiltonian (\ref{Hsemi}) onto the basis (\ref{basis}), we can obtain an effective low-energy Hamiltonian describing the surface states. For momenta in the region $-Q< k_z< Q$ the Hamiltonian reads as

\bea \label{Hedge} H_\text{edge}=\lambda k_{y}\mu^{z},\eea where $\mu^{z}$ is the Pauli matrix acting in basis $|\psi^{1,2}\rangle$. In this 2-band effective Hamiltonian for surface states, the (physical) time reversal and crystal symmetries are implemented as follows:
\bea
\Theta=\imath\mu^y\cdot\mathcal{K},~~~U_{R_z}=\imath\mu^z,~~~U_{R_y}=\imath\mu^y.
\eea
Meanwhile, the (unphysical) extra symmetries of model (\ref{H4}) acts on the surface states in the following way:
\bea
\sor=\imath\mu^z,~~~\phs=\mathcal{K}.
\eea
From this Hamiltonian it is clear that the no mass term can be added to gap out the spectrum at high-symmetry $k_z=0$ plane, due to the protection of time-reversal symmetry $\Theta$.

Away from the high-symmetry time-reversal-invariant-momenta (TRIM), time reversal symmetry is absent in a generic plane with fixed $k_z\neq0,\pi$. In such a $k_z\neq0,\pi$ plane, however, the [100] side surface states still preserves the following physical symmetries:
\bea
&\notag\tilde\Theta=U_{R_z}\cdot\Theta=\imath\sigma_x\tau_z\mathcal{K}\longrightarrow\imath\mu^x\cdot\mathcal{K},\\
&U_{R_y}=\imath\sigma_y\tau_z\longrightarrow\imath\mu^y.
\eea
as well as unphysical extra symmetries
\bea\label{sym:edge:sor}
&\sor=\imath\sigma_z\tau_z\longrightarrow\imath\mu^z,\\
&\tilde\phs=U_{R_z}\cdot\phs=\imath\sigma_y\tau_y\cdot\mathcal{K}\longrightarrow\imath\mu^z\cdot\mathcal{K}.\label{sym:edge:phs}
\eea

Although time reversal symmetry is absent, yet a combination with mirror yields a new symmetry $\tilde\Theta$. Thus, the edge states can be destroyed by adding a mass term $m_0\mu^{y}$ which respects all physical symmetries preserved on the [100] surface, as shown in \cite{KARGARIAN2016}.

Below we show that by adding the perturbations preserving all symmetries of the system in bulk, one can partially destroy the edge states, and consequently deform the double Fermi arcs to pockets as shown in Fig. \ref{fig:arcs_fermisurfaces}.

We consider two types of perturbations as follows.
\bea\label{H4:sym pertb}
H'({\bf k})=m'(\cos k_{x}-\cos k_{y})\sin k_{z}~\sigma_{z}\tau_{x},
\eea
and
\bea\label{H4:sym pertb2}
H''({\bf k})=m'' \sin k_{x}\sin k_{y}\sin k_{z}~\tau_{y}.
\eea

These perturbations preserve all physical symmetries according to transformations (\ref{symmetry1}-\ref{symmetry4}). To explore the effects of these perturbations on the edge state Hamiltonian (\ref{Hedge}), we first expand around $\mathbf{k}=(0,0,k_z)$ and keep the leading non-zero terms. Projecting the obtained Hamiltonians into the surface states (\ref{basis}), we obtain
\bea \label{pert1}
H'\propto \sin k_{z}~\mu^{y},
\eea
and
\bea \label{pert2}
H'' \propto k_{y}\sin k_{z}~\mu^{x}.
\eea
Clearly both perturbations (\ref{pert1})-(\ref{pert2}) vanish at $k_z=0$ in agreement with our previous analysis that the time reversal symmetry $\Theta$ forbids a mass term in high-symmetry TRIM plane $k_z=0,\pi$.

In a generic $k_z\neq0$ plane, both of these perturbations anti-commute with edge Dirac Hamiltonian (\ref{Hedge}). However, only perturbation (\ref{pert1}) can open a gap in the surface spectrum at $k_{z}\neq0$. Since the perturbation (\ref{pert2}) is proportional to $k_{y}$, it merely modifies the Fermi velocity and can not destroy the edge states. While a more physical understanding of this effect based on chirality mixing will be presented in the next section, here we provide a symmetry reason for this phenomena.

As shown earlier, 4-band model (\ref{H4}) exhibits two unphysical extra symmetries (\ref{sym:u1})-(\ref{sym:phs}) that are artifacts of the minimal model itself, but not an intrinsic property of the DSM materials. While the extra unitary symmetry (\ref{sym:u1}) is broken by both perturbations (\ref{H4:sym pertb})-(\ref{H4:sym pertb2}), the extra particle-hole symmetry (\ref{sym:phs}) is preserved by perturbation (\ref{H4:sym pertb2}) but not by perturbation (\ref{H4:sym pertb}). In other words, the latter perturbation Hamiltonian (\ref{H4:sym pertb2}) exhibits an extra particle-hole symmetry (\ref{sym:phs}).

As a result, if we focus on the [100] side surface states in a fixed $k_z\neq0,\pi$ plane under perturbation (\ref{pert2}), the surface states preserve the combined symmetry (\ref{sym:edge:phs}) of particle-hole operation $\phs$ and mirror reflection $U_{R_z}$. Such a symmetry $\tilde\phs=\imath\mu^z\mathcal{K}$ will forbids mass term $m_0\mu^{y}$ for edge Dirac Hamiltonian (\ref{Hedge}), while the other possible mass term $m_0\mu^x$ is forbidden by the mirror reflection symmetry $U_{R_y}$. Therefore the surface states in each fix $-Q<k_z<+Q$ plane are protected by both $\tilde\phs$ and $U_{R_y}$ symmetries, and perturbation (\ref{pert2}) cannot destroy such a double fermi arc on [100] side surface.

For the other perturbation (\ref{H4:sym pertb}) or (\ref{pert1}), on the other hand, unphysical particle-hole symmetry (\ref{sym:phs}) or (\ref{sym:edge:phs}) is explicitly broken by this perturbation Hamiltonian. As a result, no robust [100] side surface states exist in a $k_z\neq0,\pi$ plane. Therefore such a perturbation will partially destroy the double fermi arc surface states (except for TRIM plane $k_z=0$ here) and deform it into a closed fermi pocket.

\section{Weyl nodes and chirality mixing}\label{sec:chirality mixing}
The instability of the surface states can also be physically understood by the way in which the perturbations can mix the modes associated with different chiralities. In oder to understand the latter notion, it is instructive to expand the Hamiltonian (\ref{H4}) around one of the Dirac nodes, say $(0,0,Q)$ to get an insight on the low-energy excitations. The corresponding Hamiltonian reads as

\bea H_{D} =
 \begin{pmatrix}
 v_z k_{z} & 0 & 0 & vk_{-} \\
  0 & -v_z k_{z} & vk_{-}& 0 \\
  0  & vk_{+}  & v_z k_{z} & 0  \\
  vk_{+} & 0 & 0 & -v_z k_{z}
 \end{pmatrix}, \eea
where $v_{z}=t_{z}\sin Q$, $v=\lambda$, $k_{\pm}=k_{x}\pm\imath k_{y}$. This form of the Hamiltonian, as written, is not illuminating. Note that here the momentum $\mathbf{k}$ is measured from the node. We perform a unitary transformation \cite{KARGARIAN2016} to bring it to the following form.

\bea\label{DW1} \nonumber \bar{H}_D &=&
 \begin{pmatrix}
 v_z k_{z} & vk_{+} & 0 & 0 \\
  vk_{-} & -v_z k_{z} & 0 & 0 \\
  0  & 0  & v_z k_{z} & -vk_{-}  \\
  0 & 0 & -vk_{+} & -v_z k_{z}
 \end{pmatrix}\\ \nonumber \\
 &=&v_{z}k_{z}\tau_{z}+v(k_x\sigma_{z}\tau_{x}-k_y\tau_{y}), \eea
 where by abuse of notation we use $\boldsymbol{\tau}$ and $\boldsymbol{\sigma}$ to act within orbital and spin spaces in the rotated basis.

The Hamiltonian (\ref{DW1}) is a low energy description of Dirac semimetals \cite{wang:prb12,wang:prb13}. It is composed of two Weyl points located at the same momentum $(0,0,Q)$. The upper block corresponds to a Weyl node with chirality $\nu=+1$ and the lower one yields a Weyl node with opposite chirality $\nu=-1$. Thus, in the absence of coupling between Weyl nodes, the rotationally invariant disorder can not scatter the electrons between the nodes \cite{Sid:prx2014}.

The perturbations (\ref{H4:sym pertb}) and (\ref{H4:sym pertb2}) account for different high-order corrections to the Dirac Hamiltonian (\ref{DW1}). Using the same unitary transformation we can also write the latter perturbations in chiral basis. We obtain
\bea\label{DW2} \nonumber \bar{H}' &=&
 \beta_1(k_x^2-k_y^2) \begin{pmatrix}
 0 & 0 & 0 & 1 \\
  0 & 0 & 1 & 0 \\
  0  & 1  & 0 & 0  \\
  1 & 0 & 0 & 0
 \end{pmatrix}\\ \nonumber \\
 &=&\beta_1(k_x^2-k_y^2)\sigma_{x}\tau_{x}
 \eea
 for (\ref{H4:sym pertb}) and

 \bea\label{DW3} \nonumber \bar{H}'' &=&
 \beta_2 k_xk_y \begin{pmatrix}
 0 & 0 & 0 & -\imath \\
  0 & 0 & -\imath & 0 \\
  0  & \imath  & 0 & 0  \\
  \imath & 0 & 0 & 0
 \end{pmatrix} \\ \nonumber \\
 &=&\beta_2 k_xk_y \sigma_{y}\tau_{x}
  \eea
 for (\ref{H4:sym pertb2}), where we define $\beta_{1}=m'\sin Q$ and $\beta_{2}=m''\sin Q$.

These perturbations add the quadratic corrections to the Dirac Hamiltonian (\ref{DW1}). Such terms mix the chiralities and can lead to the scattering between the nodes in the presence of impurities \cite{hosur:prl12}.

One should note that mixing Weyl nodes through the perturbations (\ref{H4:sym pertb}) and (\ref{H4:sym pertb2}) will not generate a bulk gap in the spectrum, since they preserve the crystal as well as time reversal symmetries. Therefore, the presence of the perturbations have no significant effects on the bulk energy spectrum of the system. They, however, can alter the surface states in the vicinity of the projection of the nodes onto the surfaces. One such example is the [100] side surface, which is parallel to the separation between nodes in the momentum space (also parallel to the $k_{y}$-$k_z$ plane) as discussed in the preceding section.

Near the boundary of the system located at $x=0$, the momentum $k_{x}$ is no longer conserved and is replaced by $-\imath\partial_{x}$ due to the surface potential disturbing the periodicity of the lattice. As discussed the surface states decay exponentially like $e^{-x/\xi}$ into the bulk. The presence of the perturbation (\ref{H4:sym pertb}) will give rise to a mass term $H'=(\beta_{1}/\xi^2) \sigma_{x}\tau_{x}$ for the gapless surface state (\ref{Hedge}). Note that in the vicinity of the node $1/\xi\propto\delta k_{z}=k_z-Q$. Hence, the gapless surface states at $\delta k_{z}\neq0$ away from the projection of the bulk node will be destroyed, while the Dirac node remains gapless since the mass term vanishes at $\delta k_{z}=0$.

The perturbation (\ref{H4:sym pertb2}), inspite of mixing the Weyl nodes in the bulk, cannot destroy the double fermi arc surface states. In the vicinity of the surface it yields a term $H''=(\beta_{2}/\xi)k_{y}$, which vanishes along the $k_{y}=0$ line for all $k_{z}$. Therefore, although the symmetric perturbation (\ref{H4:sym pertb2}) mixes the Weyl nodes with opposite chirality, it can not deform the surface states no matter how strong it is. This is in agreement with the previous analysis, which shows that unphysical particle-hole symmetry (\ref{sym:phs}) in the 4-band minimal model protects the double fermi arc in the presence of perturbation (\ref{H4:sym pertb2}).


\section{A robust Dirac semimetal without surface states}\label{sec:dsm with no surface}

In the previous discussions, we have shown from various aspects that the double fermi arc surface states in the 4-band minimal model (\ref{H4}) of DSM material Cd$_3$As$_2$ is not robust against symmetry-preserving perturbations. In particular, there exist physical perturbations preserving all symmetries of the DSM materials, such as (\ref{H4:sym pertb}), which can continuously deform the double fermi arcs on [100] side surface into a closed fermi pocket located near surface BZ center. While an unphysical particle-hole symmetry (\ref{sym:phs}) can protect the double fermi arc surface states, it is not a physical symmetry and generically absent in a DSM material such as Cd$_3$As$_2$. However, the deformed fermi pocket near [100] side surface BZ center ($k_y=k_z=0$) is protected by time reversal symmetry $\Theta$ in TRIM $k_z=0$ plane and cannot be destroyed by any symmetric perturbation \cite{yang:ncomm2014}. Therefore in spite of the generic deformations on the double fermi arc states, the gapless surface states on [100] side surface cannot be completely removed without a bulk phase transition.

One important character for topological phases is their bulk-boundary correspondence, i.e. the one-to-one correspondence between the bulk topology and the surface states. In our case of Dirac semimetals described by model (\ref{H4}), as discussed above, both the bulk Dirac nodes and side surface states are robust against perturbations. They are indeed protected by distinct symmetries: while time reversal symmetry alone protects the [100] side surface states, stability of bulk Dirac nodes require time reversal, inversion and discrete $C_{n,z}$ rotational symmetries around (001)-axis. This already implies the unusual bulk-boundary correspondence in these Dirac semimetals, e.g. breaking $C_{n,z}$ rotation symmetry alone can gap out the bulk Dirac nodes while preserving the metallic [100] surface states, leading to a 3d TI state\cite{KARGARIAN2016}. In other words, the bulk Dirac nodes seems unrelated to the topological [100] surface states. In the following we go one step further and ask the following question: can there be a robust Dirac semimetal with no surface states at all, while keeping time reversal and all crystal symmetries?

To be concrete, we focus on the class of Dirac semimetals from band inversion \cite{Armitage2017}, which includes Cd$_3$As$_2$ and Na$_3$Bi as described in model (\ref{H4}). The Dirac nodes in these DSMs are located on high-symmetry axis of the bulk BZ (such as (001)-axis in Cd$_3$As$_2$ and Na$_3$Bi), but not at the high-symmetry TRIM points \cite{Armitage2017} such as BZ centers or corners. Below we show that the surface states can have no correspondence with bulk Dirac nodes for this type of DSMs: while preserving time reversal and all crystal symmetries, in certain DSMs the [100] side surface states can be completely removed despite the presence of stable bulk Dirac nodes.

First of all, since Dirac nodes are located at $(0,0,\pm Q)$ on the $k_x=k_y=0$ axis, there is a bulk spectrum gap in both $k_z=0$ and $k_z=\pi$ plane. Therefore one can always define
a $\mathbb{Z}_2$ topological invariant $\nu_{0(\pi)}=0,1$ in each plane $k_z=0(\pi)$. As is shown in Ref.~\onlinecite{yang:ncomm2014}, in a 4-band DSM model like (\ref{H4}) with one pair of Dirac nodes at ${\bf k}=(0,0,\pm Q)$, it is impossible to have a trivial $\mathbb{Z}_2$ invariant $\nu_{0}=\nu_{\pi}=0$ in both planes $k_z=0,\pi$. As will be shown soon, this is really an artifact of the 4-band model but not a generic properties of DSM materials. Below we write down an 8-band model of 3d DSM, which also exhibits a pair of Dirac nodes on $\hat z$-axis with the same crystal symmetries, but has no robust surface states on any open surface.

\begin{figure}
\includegraphics[width=8.5cm]{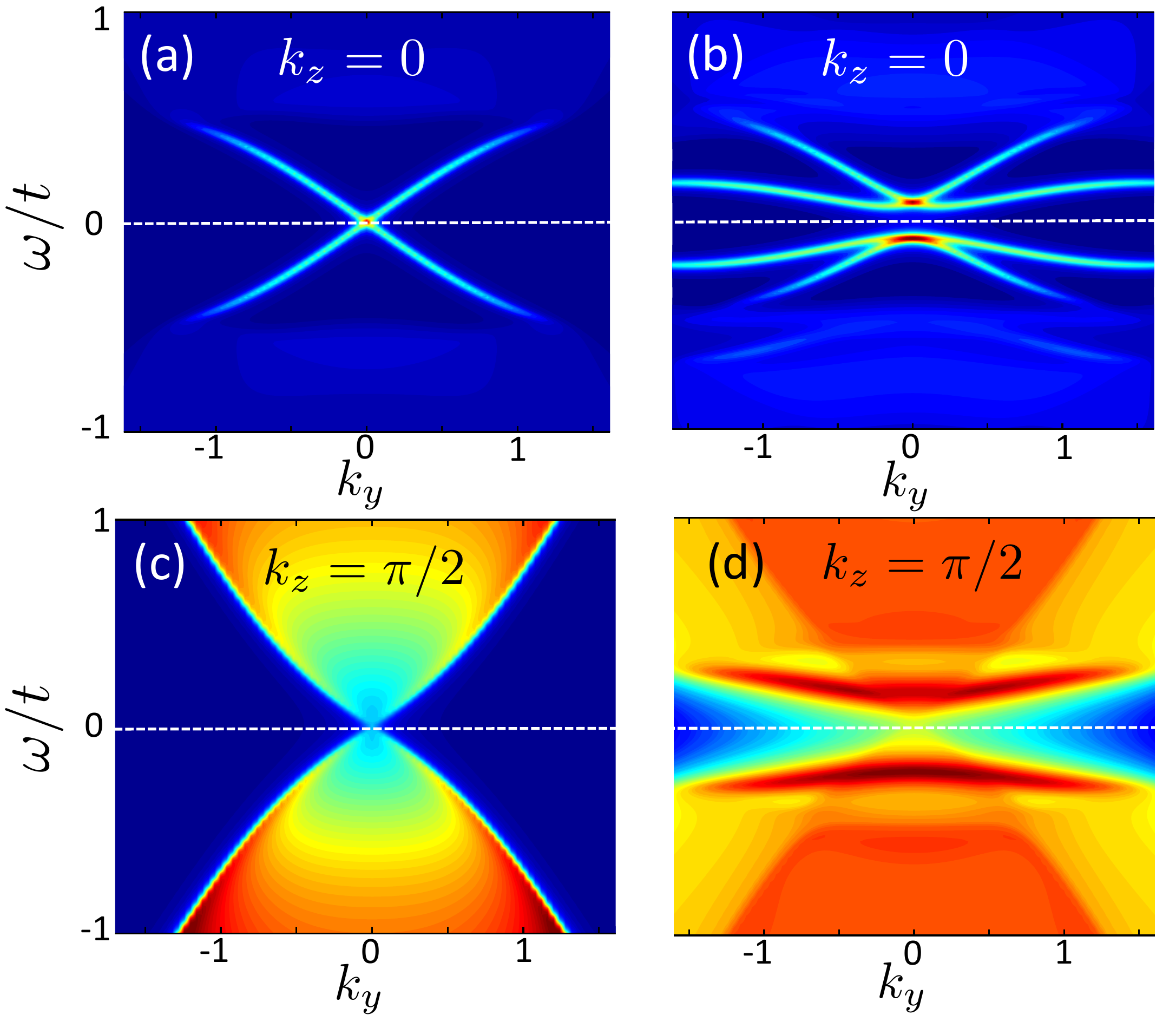}
\caption{(Color online) Spectral density of four-band (a-c) and eight-band (b-d) models at the $\omega-k_y$ plane. As shown, each panel corresponds to a fixed $k_z$. Both models include the symmetry preserving perturbation $H'(\mathbf{k})$ in (\ref{H4:sym pertb}). The dashed line corresponds to $\omega=0$. While the four-band model yields gapless surface states around $k_z=0$ and $k_z=\pi/2$, the spectrum of the eight-band model (\ref{H8}) becomes gapped around $k_z=0$ and remains gapless at $k_z=\pi/2$, where the bulk Dirac nodes are projected onto the surface.} \label{fig:8band_model}
\end{figure}

The 8-band model is as follows:
\bea\label{H8}
&H_8({\bf k})=\bpm H({\bf k})&M({\bf k})\\ M^\dagger({\bf k})&H_\text{TI}({\bf k})\epm,\\
&\notag H_\text{TI}({\bf k})=\varepsilon^\prime_{\bf k}+t^\prime(\cos k_x+\cos k_y+\cos k_z-2)\tau_z\\
&\label{H_TI}+\big[\lambda_{xy}(\sin k_x\sigma_x-\sin k_y\sigma_y)+\lambda_z\sin k_z\sigma_z\big]\tau_x,\\
&\label{mixing term} M({\bf k})=m_0(1-\tau_z),~~~m_0=\text{real}.
\eea
While time reversal symmetry is implemented as usual by $\Theta=\imath\sigma_y\cdot\mathcal{K}$, the crystal symmetries act as follows:
\bea\label{sym:c4}
&\tilde{U}_{C_{4,z}}=\bpm e^{\imath\frac\pi4\sigma_z}\imath\sigma_z\tau_z&\\&e^{-\imath\frac\pi4\sigma_z}\epm,\\
&\label{sym:c2}\tilde{U}_{C_{2,\alpha}}=-\imath\sigma_\alpha,~~~\alpha=x,y,z,\\
&\tilde{U}_I=\tau_z.\label{sym:inversion}
\eea
In particular, the upper $4\times4$ block of 8-band model (\ref{H8}) is nothing but the original 4-band model $H({\bf k})$ in (\ref{H4}), while the lower $4\times4$ block $H_\text{TI}({\bf k})$ is the minimal 4-band model of a 3d TI \cite{Qi:prb08}. The $4\times4$ matrix $M({\bf k})$ describes the mixing between the DSM block and the TI block, and it is straightforward to show that Hamiltonian (\ref{H8}) preserves time reversal and all crystal symmetries (\ref{sym:c4})-(\ref{sym:inversion}).

Below we show that the surface states of model (\ref{H8}) is generically gapped on [100] side surface. We first present a low-energy effective surface theory around $k_z=0$ plane, and show that the mixing term (\ref{mixing term}) between DSM bands and TI bands can symmetrically gap out the surface states at $k_z=0$. Then we present a full lattice calculation of the surface states in a semi-infinite geometry to demonstrate this point.

Without the mixing term $M({\bf k})$ in (\ref{mixing term}), a continuum Dirac Hamiltonian for the [100] open surface can be written as
\bea
&\notag\mathcal{H}_\text{Dirac}=(-\imath\partial_x\sigma_x+k_y\sigma_y\alpha_z)\tau_x+m(k_y=k_z=0,x)\tau_z\\
&
\eea
around $k_z=0$ plane, where $(\alpha_x,\alpha_y,\alpha_z)$ represents Pauli matrices for the DSM/TI band index. For simplicity we've set all fermi velocity as 1 and set the mass as the same for the DSM and TI blocks. A mass domain wall in $m(x)$ at $x=0$ gives rise to the zero-energy surface states
\bea
\langle x|\text{surf}\rangle\sim e^{-\int_0^x m(x^\prime)\text{d}x^\prime}|\sigma_x\tau_y=1\rangle
\eea
We use a new set of Pauli matrices to denote the low-energy subspace of surface states satisfying $\sigma_x\tau_y=1$:
\bea
(\mu_x,\mu_y,\mu_z)=(\sigma_z\tau_x\sim-\sigma_y\tau_z,\sigma_x\sim\tau_z,\sigma_y\tau_x\sim\sigma_z\tau_z)\notag
\eea
And the effective surface Hamiltonian can be written as
\bea
\mathcal{H}^0_\text{surf}\sim k_y\mu_z\alpha_z,
\eea
It's straightforward to show that the mixing term adds a mass term for the surface Dirac Hamiltonian
\bea
\mathcal{H}^1_\text{surf}=\langle\text{surf}|\bpm&M({\bf k})\\M^\dagger({\bf k})&\epm|\text{surf}\rangle\sim m_0\mu_x
\eea
which destroys the gapless surface states around $k_z=0$.

To justify the effective surface theory presented above, the spectral density $A(\mathbf{k},\omega)=\mathrm{Im}[1/(\omega-H_8(\mathbf{k})+\imath0^+)]$ of the model (\ref{H8}) in a semi-infinite geometry. Without loss of generality we assumed $\varepsilon_{\bf k}=\varepsilon^\prime_{\bf k}=0$ and other parameters used are $t_z=0.7t, \lambda=0.5t, Q=\pi/2, m'=0.8t, t'=0.5t, \lambda_{xy}=\lambda_z=0.2t, m_0=0.05t$. We focus on the [100] side surface, and the corresponding spectral densities at the $\omega-k_y$ plane are shown in Fig. \ref{fig:8band_model} for fixed values of $k_z$. For the four-band model (\ref{H4}) there are gapless surface states at $k_z=0$, panel Fig. \ref{fig:8band_model}(a), resembling the Dirac cone at the surface of a 3D TI and at the $k_z=\pm \pi/2$, panel Fig. \ref{fig:8band_model}(c). The latter points correspond to the projection of the bulk Dirac nodes onto the [100] side surface. For the eight-band model (\ref{H8}), while the projected states remain gapless due to merging to the bulk states (see Fig. \ref{fig:8band_model}(d)), the surface states are completely gaped out as shown in Fig. \ref{fig:8band_model}(b). Therefore all nontrivial surface states can be removed by symmetry-preserving perturbations.

\begin{figure}
\includegraphics[width=8cm]{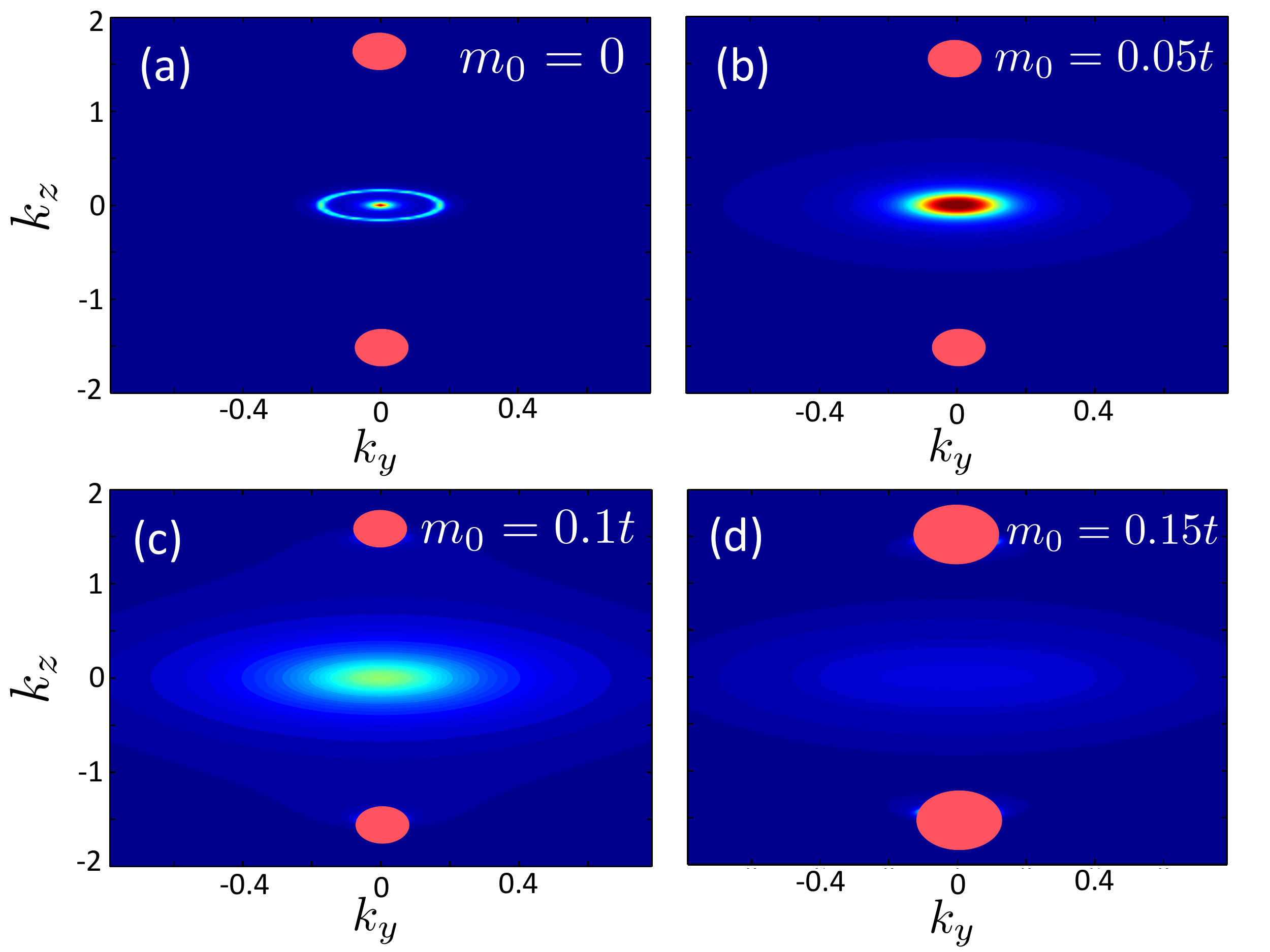}
\caption{(Color online) Spectral density $A(\mathbf{k},\omega)$ at $\omega=0$ at the surface BZ, the $k_z-k_y$ plane. Panel (a) shows decoupled systems of DSM and TI when $m_0=0$. The tip of Dirac cone at the surface of TI is located at $k_y=k_z=0$ and it is surrounded by the closed Fermi pocket of DSM. By increasing $m_0$, the coupling to the TI, the central electron pocket, resembling of the Dirac cone on the surface of TI, is completely removed from the [100] surface. The red ovals correspond to the gapless sates projected from bulk Dirac nodes.} \label{fig:kzky}
\end{figure}

In Fig. \ref{fig:kzky} we present the evolution of the surfaces states of the model (\ref{H8}) the [100] side surface. When DSM and TI parts of model are decoupled, i.e. when $m_0=0$, there are two surface states around $k_y=k_z=0$. The closed Fermi pocket (red colored) belongs to the DSM bulk and the smaller pocket located at the origin is due to the surface states of TI. Note that  for clarity we have taken $\varepsilon_{\mathbf{k}}\neq0$ and $\varepsilon_{\mathbf{k}}'=0$, pinning the Fermi level at the surface Dirac cone of TI. As seen in Fig. \ref{fig:kzky} (b-d) the coupling between DSM and TI hybridizes the Fermi pockets and completely removes the surfaces states at $k_z=0$.

While this example illustrates that there is no correspondence between bulk Dirac nodes and protected surface states in a DSM from band inversion, there is another bulk-boundary correspondence that dictates the robustness of surface states. To be concrete, for a DSM from band inversion where Dirac nodes are located on high-symmetry axis (but not at high-symmetry points such as zone center and zone corner), one can define a $\mathbb{Z}_2$ topological invariant despite the gapless Dirac points in the bulk. Since the spectrum remains gapped at the 8 time reversal invariant momenta (TRIM) of the 3d BZ, the so-called Fu-Kane invariant is well-defined even for a DSM which preserves both inversion and time reversal symmetries:
\cite{Fu:prb07}
\bea (-)^{\nu}=\prod_{i}\delta_i, ~~~\delta_{i}=\prod_{m=1}^{N}\xi_{2m}(\Gamma_i), \eea
Here $\xi_{2m}=\pm1$ is parity eigenvalue of the $2m$-th occupied band among all $2N$ filled bands, and $\{\Gamma_{i}\}$ represent the TRIM. We found that for four-band model $\nu=1$, corresponding to the robust topological surface states. On the other hand, $\nu=0$ for eight-band model with no robust surface states. This can be viewed as a generalization of the usual bulk-boundary correspondence to the DSMs from band inversions.

\section{Conclusions}\label{sec:conclude}

Motivated by the discovery of DSM materials, in this paper we address two issues regarding the stability and deformation of surface states in a DSM from band inversions.

Focusing on a minimal model for DSM materials Cd$_3$As$_2$ and Na$_3$Bi, we show that the double fermi arcs in the minimal model can be stablized by a non-generic particle-hole symmetry, which is absent in a generic material. We further show that physical perturbations without this particle-hole symmetry can mix the two Weyl fermions with opposite chiralities away from the Dirac points, which continuously deforms the surface double fermi arcs into a closed fermi pocket without destroying the bulk Dirac nodes.

We also show in a DSM from band inversions, there is an unusual bulk-boundary correspondence between bulk topology and surface states. In particular, the bulk Dirac points have no correspondence to the surface states, while another Fu-Kane $Z_2$ invariant well-defined in the bulk determines the stability of surface states. We explicitly construct an 8-band model for such a DSM, with exactly the same crystalline symmetries as Cd$_3$As$_2$, supporting no surface states at all in spite of the bulk Dirac points. We show that this is related to the bulk $Z_2$ topological invariant, which can be defined even in the presence of the bulk Dirac points. It will be interesting to look for a material candidate for such a DSM with no surface states.

\acknowledgments MK acknowledges the Sharif University of Technology's office of Vice President for Research. YML and MR acknowledge the support of the Center for Emergent Materials, an NSF MRSEC, under award number DMR-1420451.

%

\end{document}